\documentclass{aa}
\usepackage{graphicx}

\usepackage{times}
\usepackage{epsfig}
\usepackage{epsf}
\usepackage{rotating}

\begin{document}


\title{{\it GALEX} high time-resolution ultraviolet observations of dMe flare events}


\author{
B.Y. Welsh,\inst{1}
J. Wheatley,\inst{1}
S.E. Browne,\inst{1}
O.H.W. Siegmund,\inst{1}
J.G. Doyle,\inst{2}
E. O'Shea,\inst{2}
A. Antonova,\inst{2}
K. Forster,\inst{3}
M. Seibert,\inst{3}
P. Morrissey \inst{3}
and Y. Taroyan \inst{4}}


\date{Submitted: 30/03 2006}

\titlerunning{$\it GALEX$ dMe flare events}
\authorrunning{Welsh et al.}


\abstract
{}
{We present near ultraviolet (NUV:1750 - 2800\AA) and far ultraviolet
(FUV: 1350 - 1750\AA) light-curves for flares on 4 nearby dMe-type stars (GJ 
3685A, CR Dra, AF Psc and SDSS J084425.9+513830.5) observed with the $\it 
GALEX$ satellite.}
{Taking advantage of the time-tagged events recorded with the $\it GALEX$ 
photon counting detectors, we present high temporal resolution ($<$ 0.01 sec) 
analysis of these UV flare data.}
{A statistical analysis of 700 seconds of pre-flare quiescence data for both 
CR Dra and SDSS J084425.9+513830.5 failed to reveal the presence of 
significant micro-flare activity in time bins of 0.2, 1 and 10 second 
intervals. Using an appropriate differential emission measure for both the 
quiescent and flaring state, it is possible to reproduce the observed FUV:NUV 
flux ratios. A major determinant in reproducing this flux ratio is found to be 
the value of plasma electron density during the flare. We also searched 
the count rate data recorded during each of the four flare events for 
periodicity associated with magneto-hydrodynamic oscillations in the active  
region coronal loops. Significant oscillations were detected during the 
flare events observed on all 4 stars, with
periodicities found in the 30 to 40 second range. Flare oscillations with this 
periodicity can be explained as acoustic waves in a
coronal loop of length of $\approx 10^{9}$ cm for an assumed plasma 
temperature of $5-20 \times 10^{6}$K. This suggests a loop length for these 
M-dwarf flares of less than $1/10^{th}$ of the stellar radii. We believe that 
this is the first detection of non-solar coronal loop flare oscillations 
observed at ultraviolet wavelengths.}
{}



\maketitle


\section{Introduction}
Although flare stars were detected early in the last century, it was not until
the late 1940's that flare star research attracted
wide-spread attention with the observation by Joy $\&$
Humason (\cite{joy49}) of a four magnitude eruptive event. Since then there have been many
hundreds of publications concerning these active stellar sources. The first coincidence
between optical and radio
activity was reported by Lovell $\&$Solomon (\cite{Lovell66}), with the first X-ray flare being
observed by Heise et al. (\cite{Heise75}). Surveys show that low mass M-dwarf (dMe) stars 
account for more than 75\% of the stellar 
population in the solar neighborhood and, due to the strong magnetic fields 
that cover most of their stellar disks, many of them exhibit significant levels 
of coronal activity. In a recent survey of 8000 late-type dwarfs contained in 
the Sloan Digital Sky Survey (SDSS) catalog, West et al. (\cite{west04}) have found
that $>$ 50\% of stars of spectral class M4 to M9 have high levels
of magnetic activity, as inferred by the presence of the H$\alpha$ emission 
line in their stellar spectra. The most easily observable manifestation of this
activity is that of stellar flare eruptions, as typified by the star UV Ceti, 
which involves random outbursts in which the short-term stellar brightness 
increases significantly on time-scales of seconds to hours at 
X-ray, UV, visible and radio wavelengths (Schmitt et al. \cite{schmitt93}; 
Phillips et al. \cite{phillips88}; Stepanov et al.
\cite{stepanov95}).
Although the detailed physics associated with the formation of stellar flares 
is still uncertain (Haisch, Strong $\&$ Redono  \cite{haisch91}), it is clear that these eruptions are 
linked to the magnetic heating processes occurring in stellar coronae. We note 
that the soft X-ray emission from flares (which is indicative of gas with 
temperatures of $>$ 10$^{6}$K) is dominated by coronal line-emission, while 
the UV radiation during flare events originates in the chromosphere and 
transition region and is characterized by emission from a gas with a  
temperature of $\sim$ $10^{5}$K. There is still much debate as
to the relative contributions from emission line and/or UV continuum flux to the
total heating budget in flare stars (Robinson et al. \cite{robinson01}; Hawley et 
al.\cite{hawley03}; G\"{u}del et al.\cite{gudel03}).
\begin{table*}
\begin{center}
\caption{Summary of $\it GALEX$ flare star observational parameters}

\begin{tabular}{lllccc}
\hline
\hline
Star Name&R.A.(J2000)&Dec.(J2000)& Survey Mode&MAST Identifier&Exposure (sec)\\
\hline
GJ 3685A&11:47:41&+00:15:12&MIS&MISDR1-13062-0283-0003&1244\\
AF Psc&23:31:44&-02:45:12&GI*&GI1-067026-PGC71626-0002&1705\\
CR Dra&16:17:05&+55:16:09&DIS&ELAISN1-07-0004&1650\\
SDSS J084425&08:44:26&+51:38:31&MIS&MISDR1-03333-0447-0001&1669\\
\hline
\multicolumn{6}{l}{* pointed Guest Investigator observation} \\
\hline
\hline
\end{tabular}
\end{center}
\end{table*}
Recent EUV and X-ray observations suggest that flares statistically contribute 
a significant fraction of the overall (X-ray) heating of the coronae on all 
magnetically active stars (G\"{u}del et al. \cite{gudel03};
Arzner $\&$ G\"{u}del \cite{arzner04}).
In particular, the recent 
$\it XMM-Newton$ satellite observations of dMe star flares by
Mitra-Kraev et al. (\cite{mitra05a}) 
have noted a correlation between the observed ultraviolet (UV) flare 
energy and the corresponding increase in X-ray luminosity, with the
UV flares temporally preceding the peaks in X-ray flux. These observations
would seem to support the predictions for the Neupert Effect
(Neupert \cite{neupert68}), as outlined in the chromospheric
evaporation model for solar flares (Antonucci,
Gabriel $\&$ Dennis \cite{anton84}).
In this picture, during the impulsive phase of
a stellar flare, the UV/optical emission due to
accelerated electrons that gyrate downward
along the magnetic field lines and impact the chromosphere,
should precede the more slowly evolving X-rays that are emitted by the heated
plasma. 

Although flare events have been routinely recorded on many M-dwarfs at 
radio(Jackson, Kundu $\&$ White \cite{jackson89}), optical (Gunn
et al. \cite{gunn94}), Ultra-violet (Robinson et al. \cite{robinson01}), Extreme
Ultra-violet (Audard et al. \cite{audard00})
and X-ray (Marino, Micela $\&$ Peres \cite{marino00}) wavelengths, such observations have 
generally suffered from both poor temporal-resolution and the inability to 
track a large event throughout the entire pre-flare to post-flare period. 
However, Beskin et al. (\cite{beskin88}) recorded emission structure on time-scales of 0.3 
-- 0.8 seconds in high-temporal (0.3 milli-second) U-band visible observations of 
flare events on several bright UV Ceti-type stars. Using the scale height of
the atmosphere and the velocity of the shock led them to the 
conclusion that optical flares on M-dwarfs were due to thermal phenomena.  
We note that U-band observations of both solar and stellar
flares have been widely used as a proxy for
non-thermal hard X-ray 
flare emission (Neidig \cite{neidig89}, Hawley et al. \cite{hawley95}).

High ($<$ 1 sec) 
temporal resolution observations of ultraviolet flares can provide detailed 
information on microflaring activity and also determine whether wave activity is present. Both
of these are thought to be important in the heating of coronae 
(Robinson, Carpenter $\&$ Percival \cite{robinson99};  Roberts \cite{roberts04}).
Furthermore, the ability to record 
emission throughout an entire energetic flare event can provide insights into 
the nature of the pre-flare and post-flare quiescent phases and how their 
statistical small-scale flux variations may be related to the onset of a major 
flare event (G\"{u}del et al. \cite{gudel04}). The NASA $\it GALEX$
satellite (Martin et al. \cite{mar05}), 
launched in 2003, has recently been shown to be an excellent observational 
platform for the serendipitous detection and high time-resolved observation
of stellar flares in two ultraviolet (UV) photometric bands. For example, an 
extremely large flare event was observed on the dM4e star GJ 3685A on April 
20th, 2004 that revealed an increase in stellar brightness of $>$ 10 UV 
magnitudes in a period of $<$ 200 seconds (Robsinson et al. \cite{robinson05}). These UV 
observations provided compelling evidence for two distinct classes of flares 
during this eruption, each characterized by significantly different ratios of 
their measured near UV (NUV) and far UV (FUV) fluxes. 
In this Paper we re-visit those data (which
were analyzed under the assumption of a blackbody
energy distribution) and in addition, we
report on three further flare 
events on the stars CR Dra, AF Psc and SDSS J084425.9+513830.5
(hereafter referred to as SDSS J084425) that were serendipitously 
recorded with a high time resolution of $\sim$ 0.01 sec by the $\it GALEX$ 
satellite during its all-sky survey observations during the period 2004 - 2005. 
\begin{figure*}[hbt]
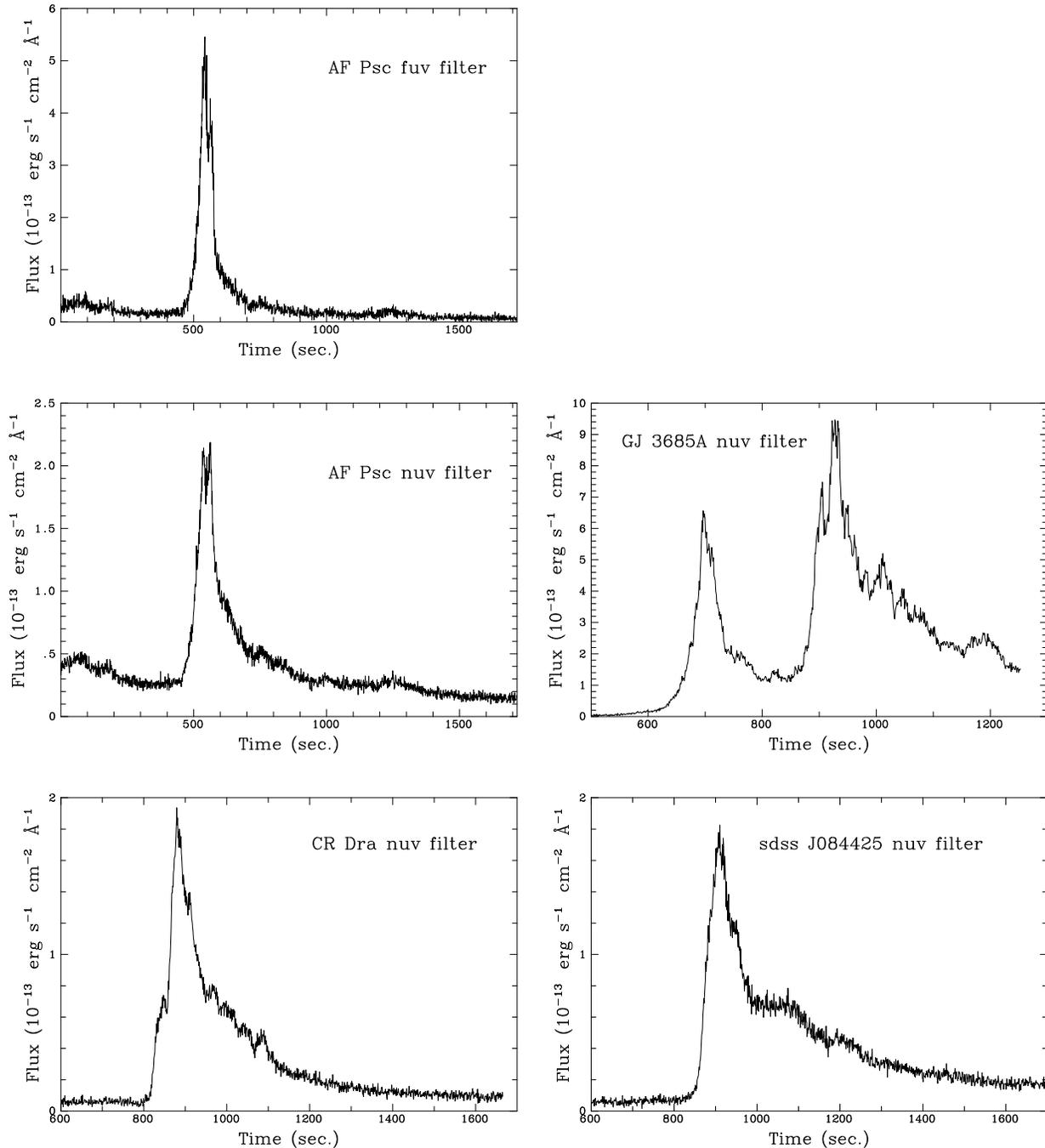

\vspace{18cm}
\includegraphics{Fig1a.ps}
\includegraphics{Fig1b.ps}
\includegraphics{Fig1c.ps}
\includegraphics{Fig1d.ps}
\includegraphics{Fig1e.ps}
\includegraphics{Fig1f.ps}

\vspace*{-0.5cm}

\caption{The calibrated flare light-curves for AF Psc, CR Dra, GJ 3685A and SDSS
J084425 for both the FUV and NUV bands.}
\label{flares}
\end{figure*}

\begin{figure*}[hbt]
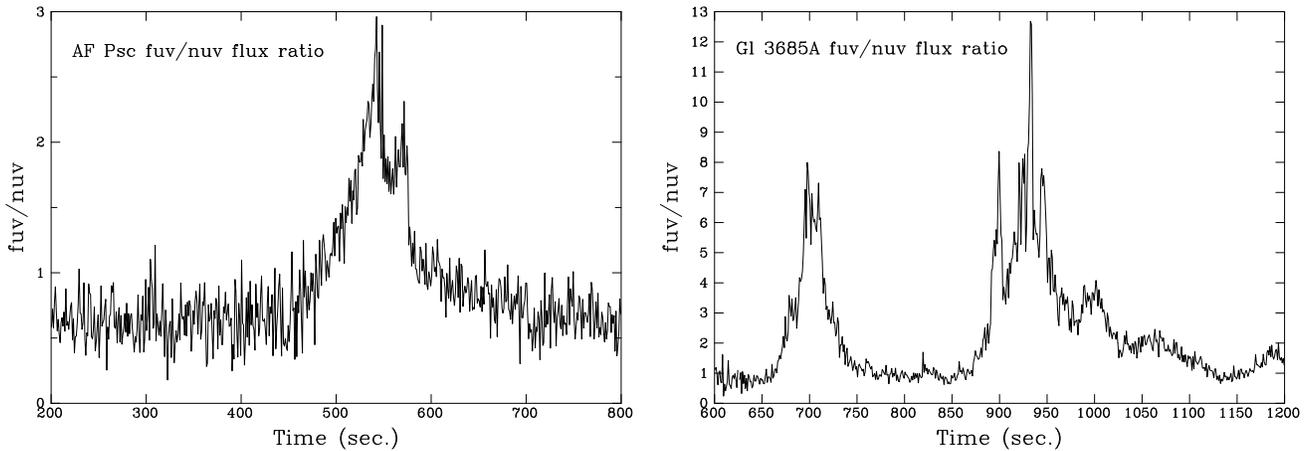

\vspace{7cm}
\includegraphics{Fig2a.ps}
\includegraphics{Fig2b.ps}

\vspace*{-1cm}

\caption{The $\it GALEX$ FUV:NUV flux ratios for the two flare stars AF Psc and GJ 3685A.}
\label{flares}
\end{figure*}

\begin{figure*}[hbt]
\vspace{17cm}
\includegraphics{Fig3a.eps}
\includegraphics{Fig3b.eps}

\vspace*{-2.5cm}

\caption{Pre-flare quiescent photon data for CR Dra (left) and SDSS J0084425 (right) binned in 0.2, 1.0 and 10.0 second intervals, shown together with the 3-$\sigma$ statistical error limits (dashed lines). No evidence is found for micro-flare activity.}
\label{flares}
\end{figure*}
\section{Observations and Data Reduction}
All of the flare events reported in Table 1 were serendipitously recorded 
within the 1.2$^{\circ}$ field of view of the $\it GALEX$ telescope in the 
far UV (FUV: 1350 - 1750\AA) and/or near UV (NUV: 1750 - 2800\AA) photometric
bands during imaging survey observations by the satellite (see Martin et al. \cite{mar05}
for a detailed description of the $\it GALEX$ instrument and its operational 
modes). These observations are listed as either Medium Imaging Survey (MIS) or 
Deep Imaging Survey (DIS) modes in Table 1, together with their total on-orbit 
exposure times and their data-file identifier as listed in
the Multi-Mission Archive at the Space Telescope
Science Institute (MAST). The detection of the flare on the 
star AF Psc was recorded during a pointed observation of a star-field selected 
for an observation of a galaxy under the auspices of NASA's $\it GALEX$ Guest 
Investigator (GI) program.

The FUV and/or NUV data for each flare were recorded as time-tagged events (with a 
time resolution of $\sim$ 0.01 sec) using the $\it GALEX$ photon counting 
detectors (Jelinsky et al. \cite{jelinsky03}). These observations have been processed (as a 
non-standard data-product) using Version 4.0 of the $\it GALEX$ Data Analysis 
Pipeline operated at the Caltech Science Operations Center (Pasadena, CA), 
which has been described in detail by Morrissey et al. (\cite{mor05}). The final data product is a 
flat-field corrected photometric time sequence of photons mapped in Right Ascension and 
Declination to the sky. 
Photon events associated with a stellar flare were summed within a 10 
arcsec aperture centered at the expected position of the associated dMe star, while 
photons associated with the background emission were summed within an annulus 
extending from $\sim$ 15 to 20 arcsec from the central position of the source. 
We assumed that any (faint) astronomical sources contained within the background
annulus remained constant during the bright flare events. 

The photon count rates associated with each of the four flare events have been converted into
UV fluxes of erg cm$^{-2}$ s$^{-1}$ \AA$^{-1}$ using the appropriate $\it GALEX$
FUV and NUV-band calibrations listed in Morrissey et al. (\cite{mor05}).
 Using these flux corrected data we 
show the FUV and NUV light-curves as a function of elapsed time for the four flare 
events in Figure 1 (we note that only NUV data was
recorded for the flares detected on CR Dra and SDSS J084425 due
to the FUV detector being turned off for satellite operational reasons).
In Figure 2 we also
show the ratio 
of the measured FUV to NUV fluxes as a function of elapsed time for both
AF Psc and GJ 3685A, which (as 
described later) can be used as a probe of the temperature variation 
throughout the duration of these flares.

A general comparison of the light-curves shown in Figure 1 reveals that in all   
cases there was a very steep and rapid rise in flux (in both the NUV and FUV
bands) over a time-scale of $\sim$ 70 seconds. Following this
flare peak intensity all the flares exhibit an 
exponential-like decay that lasts for several
hundred seconds. In the case of GJ 3685A there are multiple, smaller intensity 
flares during this decay phase. The smaller FUV flare on GJ 3685A, which is the least 
energetic of the events we have detected, is somewhat anomalous in that it 
seems to be composed of several closely spaced (in time) small flares that 
gradually increase in intensity over a time-span of $\sim$ 40 seconds.
The peak FUV flux then dies away very quickly to the pre-flare basal intensity in 
less than 50 seconds. Thus the character of the small flare on
GJ 3685A exhibits an almost inverse behavior to the larger FUV
and NUV flares recorded on the other stars, which generally consist of a
very rapid increase in flux followed by an extended exponential-like flux decrease 
that is often associated with small flares prior to
returning to the pre-flare background intensity. We also note that the NUV flare on CR
Dra, which exhibits several small brightenings during its decay phase,
does not exactly temporally track the FUV behavior, suggesting 
an origin in a slightly different temperature region.

\section{Data Analysis}

\subsection{Micro-flare activity}
Lu $\&$ Hamilton (\cite{lu91}) suggested that solar flares consists of an avalanche of many very 
small reconnection events termed micro-flares (or nano-flares). These 
small-scale phenomena are also thought to be of fundamental importance in the 
heating of the solar (and stellar) coronae (Parker  \cite{parker88};
Doyle $\&$ Butler \cite{Doyle85}). The 
statistical analysis of rapid time-sequence 
UV observations of flare activity on two dMe stars by Robinson et al. 
(\cite{robinson95}; \cite{robinson99}) 
has shown that over a time-scale of a few hours most
of the observed increase in integrated flux arises in a slowly varying stellar 
background whose short-term variability seems consistent with the presence of 
many overlapping micro-flare events. In the case of the very large flare 
observed by $\it GALEX$ on GJ 3685A, the stellar UV flux increased by a factor 
of 5 in the pre-flare period prior to the main flare event (Robinson et al. \cite{robinson05}). 
However, a statistical analysis of the pre-flare activity of this energetic 
eruption failed to reveal compelling evidence for an increased level of 
micro-flaring events.

Due to the relatively short duration ($\sim$ 1500 sec) of each
flare observation by $\it GALEX$, the 
present time-tagged photon data are not ideally suited for a rigorous
statistical investigation of micro-flare activity, which is
best performed
over far longer periods (Robinson et al. \cite{robinson95},
\cite{robinson99}). However,
for the 700 seconds of `quiescent' pre-flare NUV data for CR Dra and
SDSS J084425 we have performed a simple
test of the statistical significance of the variability in the recorded 
photon flux (Robinson et al. \cite{robinson95}). Briefly, this method involves 
searching the pre-flare photon data for time periods during which the count rate deviates 
significantly above that expected from Poisson statistics. Large micro-flares 
reveal themselves if the data is binned into small time intervals ($<$ 1 sec), 
whereas weak micro-flare events require binning over far longer periods ($\sim$ 10 sec). A 
micro-flare is deemed a real event if it is detected with a probability  
$<$ 10$^{-2}$ of it occurring by chance (i.e. the microflare event is $>$ 3$\sigma$
above the statistical mean of the data).

In Figure 3 we show 700 seconds of the pre-flare photon data for CR Dra and SDSS J0084425
binned into 0.2, 1.0 and 10.0 second intervals together with the statistical mean
levels and their
respective $\pm$3$\sigma$ limits. Unfortunately none of these plots reveal the presence
of a significant flux contribution from any major micro-flare activity. However, due to the limited
period of pre-flare observation
our present findings do not
preclude the presence of micro-flaring immediately prior
to the beginning of our observations.

\section{Results}
\subsection{The UV Flare Spectrum}
The $\it GALEX$ far ultra-violet band response extends from $\sim$1350\AA\ to 
1750\AA\ while the near ultra-violet band extends from $\sim$1750\AA\ to 
2800\AA. As already seen from $\it IUE$, $\it FUSE$ and $\it HST$ spectra of 
active dMe stars, these spectral windows contain many emission lines. Based on 
data from the above science missions, the differential emission measure (DEM) 
distribution of dMe stars has been shown to be similar to the Sun. However, the 
lack of simultaneous observations over a sufficiently wide wavelength range has 
resulted in DEM distributions derived over limited temperature intervals 
(del Zanna et al. \cite{delzanna02}, Osten et al. \cite{Osten05}). 
In the present work we use the volume emission measure of Osten et
al. (\cite{Osten05}) as a starting
point for the active dMe star EV Lac, and then convert it to the differential
emission measure distribution  (for a discussion of the
different definitions see del Zanna et al. \cite{delzanna02}). 
Using the CHIANTI package (Landi et al. \cite{Landi06}) and the EV
Lac DEM, we have derived the emission line spectrum over the wavelength interval 1200
to 3000~\AA. Multiplying by the FUV and NUV response functions then gives us the
emission lines contained within the two filters. We note
that the continuum contribution from
two--photon, bremsstrahlung and free-bound processes are $\it all$ included in these
calculations. The FUV and NUV filter contributions were calculated for a range
of electron pressures from $10^{15}$ to $10^{19}$~cm$^{-3}$~K. For pressures
around $10^{16} - 10^{17}$~cm$^{-3}$~K, the resulting FUV:NUV filter ratios
were around 0.2, substantially less than the observed quiescent value of $\sim$0.6
for AF Psc and $\sim$1 for GJ 3685A.  
\begin{figure}[hb!]
\vspace{6.5cm}
\includegraphics{Fig4.ps}

\label{fg:DEM}
\caption{Sample DEM's for a dMe star in quiescence (Q) and flaring state (F) (see text)}
\end{figure}

As mentioned above, several temperature intervals within the DEM distribution
are poorly constrained, e.g. the upper chromosphere/lower transition region.
Reducing the EV Lac DEM by a factor of five in the interval log T$_e$ = 4.0 to
4.6 results in a FUV:NUV ratio of 0.5 to 1.0. Furthermore, reducing the coronal
DEM contribution around $2 \times 10^6$~K by a factor of 3--4 gives FUV:NUV ratios close to
unity even at low electron pressures. In Table 2 we tabulate the FUV:NUV filter
ratios for such an amended quiescent DEM distribution as shown in Figure 4.

{\bf For the flaring DEM distribution we increased the emission measure over all temperatures
ranges, in particular in the region around 100,000~K where the
important C~{\sc iv} line is formed. In the
third column of Table 2 we list the FUV:NUV filter ratios, here we see values above 3
for the high electron pressure. This shows that in addition to changing the DEM
distribution, the electron pressure plays an important role. 
In earlier work on this data, Robinson et 
al. (\cite{robinson05}) used blackbody radiation
emission in an attempt to  reproduce the observed 
measured FUV:NUV ratios, but found that an additional source of
emission was required which he suggested was line emission. With a suitable 
DEM distribution we derive FUV:NUV ratios ranging from 0.59 to 0.93 for 
electron pressures of $10^{15} -- 10^{16}$ cm$^{-3}$K in excellent agreement 
with the observed pre-flare values for AF Psc and GJ 3685A. Using the flare 
DEM and a higher electron pressure gives FUV:NUV ratios around 3, in good 
agreement with the AF Psc flare but a factor of two less than that observed 
in the GJ 3685A flare. Combining the above emission line contribution with a possible 
``white light'' contribution  resulting from a higher temperature than the 
star's photosphere could lead to larger FUV:NUV for the more energetic 
flares. Whether this ``white light'' contribution is important 
in the far UV remains uncertain.

Regarding other emission contribution possibilities, previous studies of UV and X-ray emission 
in  solar flares have shown an excellent temporal correlation (Cheng et al. \cite{cheng81}),
indicating that they might both be a result of the same process of 
bombardment of the atmosphere by high energy particle beams. 
However, more recent data (Alexander \& Coyner \cite{alex06}) caution against such a 
preconception. These  authors found that the UV and hard X-ray sources were 
spatially separated, and although they showed a high temporal correlation there
was evidence for a magnetic connection between the two regions. Whether this 
could enhance the UV continuum emission during dMe flares is unclear. 

A more likely continuum enhancement mechanism is silicon reconnection. Phillips 
et al. (\cite{phillips92})  discussed the continuum observed in the  far-ultraviolet 
region from IUE spectra of dMe stars in a flaring state compared with 
that observed in solar flares. They suggested that the excitation of this 
continuum was due to the ionization of neutral silicon atoms near the 
temperature minimum region irradiated by ultraviolet line radiation emitted 
by the upper chromosphere or transition region. This also appears to be the 
case for two flares on the RS CVn star (II Peg), and one on the spotted 
active binary BY Dra. The evidence is an observed proportionality of the 
continuum intensity with the intensities of strong transition region lines 
such as C~{\sc iv}~1548.51\AA\ and C~{\sc ii}~1335.36\AA. This process is not 
presently included in CHIANTI and will affect the FUV filter at wavelengths 
less than 1682\AA.}

Using our presently proposed flare DEM at an electron pressure of $10^{18}$~cm$^{-3}$~K, the 
dominant emitter is C~{\sc iv}, contributing 35\% 
of the FUV filter with Si~{\sc iv} contributing another 8\%. 
The contribution from weaker lines and the continuum are not insignificant, e.g. 
between 1560~\AA\ and 1750~\AA, there is $\sim$27\% contribution. 
In the NUV filter, the dominant line emitters (originating mostly in the
upper chromosphere/lower transition region) are  Mg~{\sc ii}, Fe~{\sc ii},
Al~{\sc iii}, C~{\sc iii}, etc. For example, Mg~{\sc ii} has $\sim$ 10\%
contribution, the region from 2320 to 2530~\AA\ is $\sim$17\% (mostly Fe~{\sc
ii}), while the lines around 2290 $\pm$ 20~\AA\ supplies $\sim$14\%. The remaining
NUV flux contribution arises from the various continuum processes.

Several coronal lines are also present in both filters, e.g. Fe~{\sc xxi}~1354,
Fe~{\sc xi}~1467, O~{\sc vii}~1623, Fe~{\sc xxi}~2298, Fe~{\sc xx}~2666 and Ca~{\sc
xvi}~2731. The accumulated contribution of these coronal lines in the FUV filter 
is around 2\%, rising to over 10\% in the NUV filter.

These calculations therefore suggest that changing the DEM distribution can 
change the FUV:NUV flux ratio by a factor of 2--3. However, a major influence 
on the ratio concerns the increase in electron pressure during the flare.
Obviously, with a different DEM these derived emission fractions will change.  
In Figure 5 we show the range of possible spectral lines present in the 
$\it GALEX$ FUV and NUV bands during flares on dMe stars based on the DEM (F) 
distribution from Figure 4.  

Very few M dwarfs have directly measured electron densities in the transition
region or corona. Testa et al. (\cite{Testa04}), Ness et al. (\cite{Ness04}) and Osten et al.
(\cite{Osten05}) give electron densities measurements for a few dMe stars including EV
Lac. Using the O~{\sc vii} lines formed around $2.5 \times 10^6$~K, Testa et al. (\cite{Testa04})
give 
N$_e$ $\sim$ $5.6 \times 10^{10}$~cm$^{-3}$, rising to $5.6 \times 
10^{12}$~cm$^{-3}$ from Mg~{\sc xi} formed around $8 \times 10^6$~K. Values given by 
Osten et al. are slightly larger, while those given by Ness et al. are similar 
at $2.5 \times 10^6$~K rising to just over $10^{11}$~cm$^{-3}$ at $4 \times 10^6$~K as derived 
from Ne~{\sc ix}. At lower temperatures, e.g. $2.5 \times 10^5$~K, Osten et al. derived an electron
density of $\sim$ $10^{11}$~cm$^{-3}$ from O~{\sc v}. This suggests electron
pressures ranging from $\sim$ 3$ \times 10^{16}$~cm$^{-3}$~K at $2.5 \times 10^5$~K to
$\sim$ 4$\times 10^{19}$~cm$^{-3}$~K at $8 \times 10^6$~K. It should however be noted that the derived
density at these higher temperatures are a little uncertain due to line blending problems. As an
alternative to the contact pressure assumption, we re-run the models for a constant electron
density of $10^{11}$~cm$^{-3}$ and $10^{12}$~cm$^{-3}$. For the DEM (F) distribution, this gives
an FUV:NUV flux ratio of around 2.5.

\begin{figure*}[hbt]
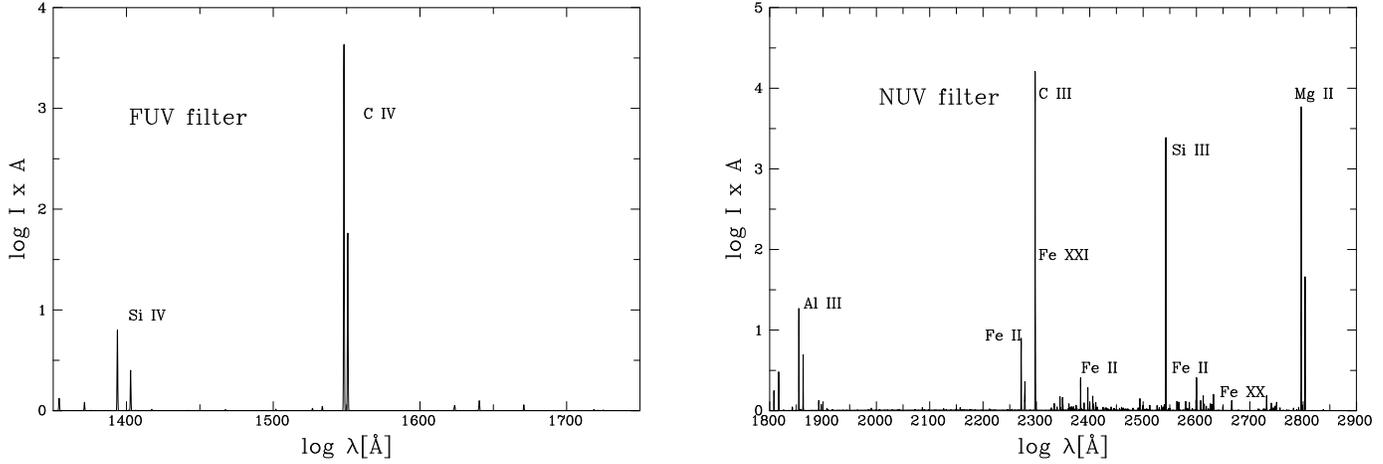

\vspace{7cm}
\includegraphics{Fig5a.ps}
\includegraphics{Fig5b.ps}

\vspace*{-1.0cm}

\caption{Sample dMe star flare emission spectrum with line identifications after 
folding the spectrum derived from the (F) DEM in Figure 4 with an electron pressure of
$10^{18}$ cm$^{-3}$~K through the $\it GALEX$ FUV and NUV band response functions 
(see text).}
\label{flares}
\end{figure*}

\begin{figure*}[hbt]
\vspace{10.0cm}
\includegraphics{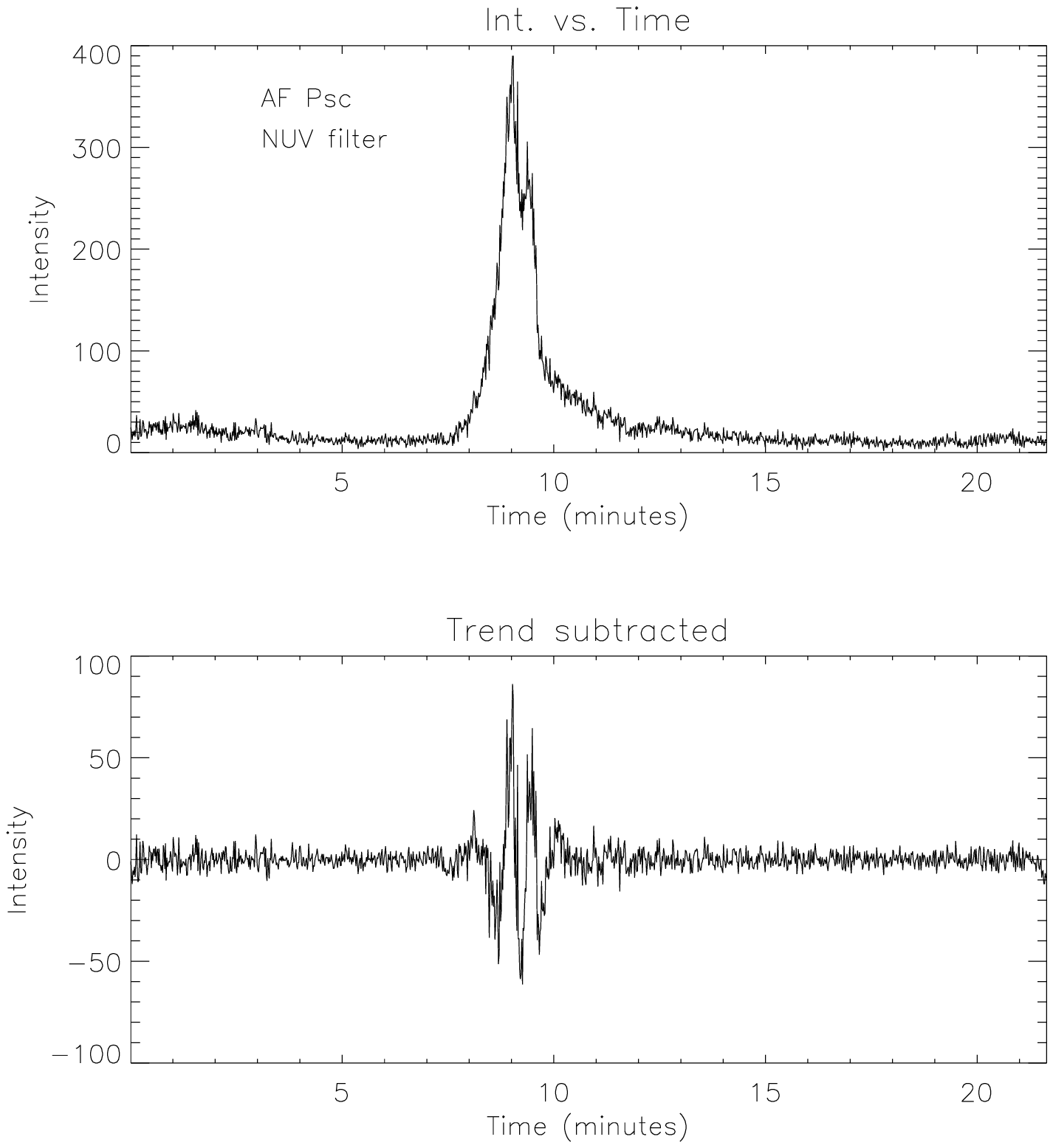}
\includegraphics{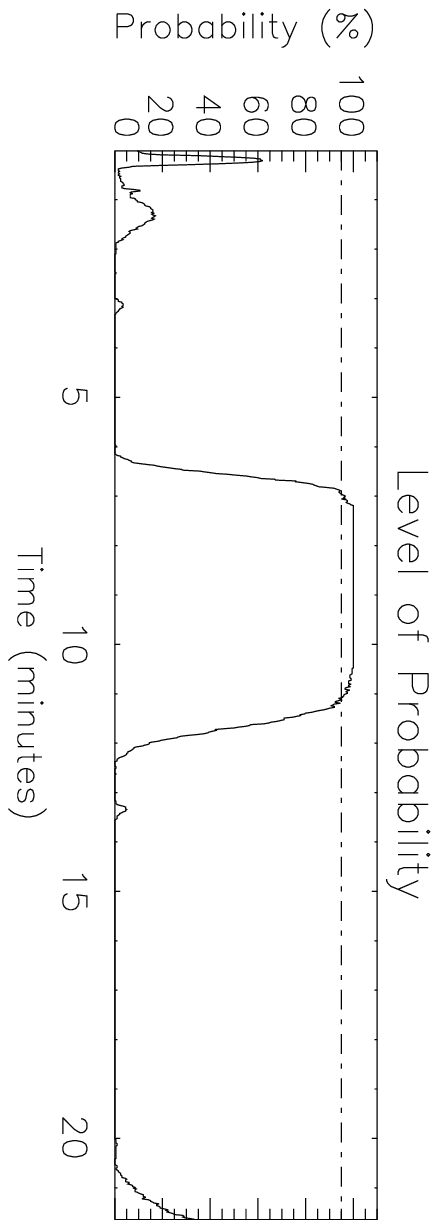}
\includegraphics{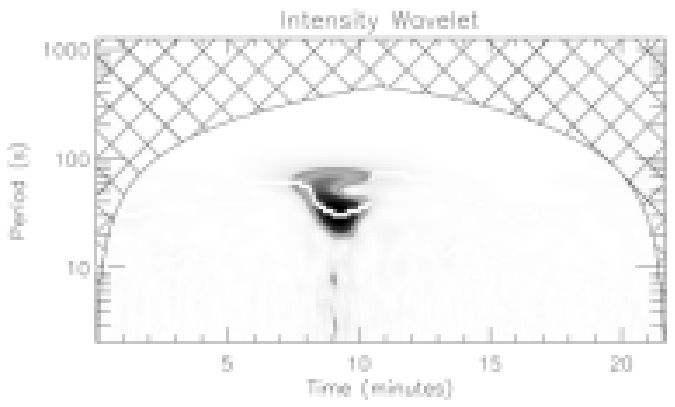}
\includegraphics{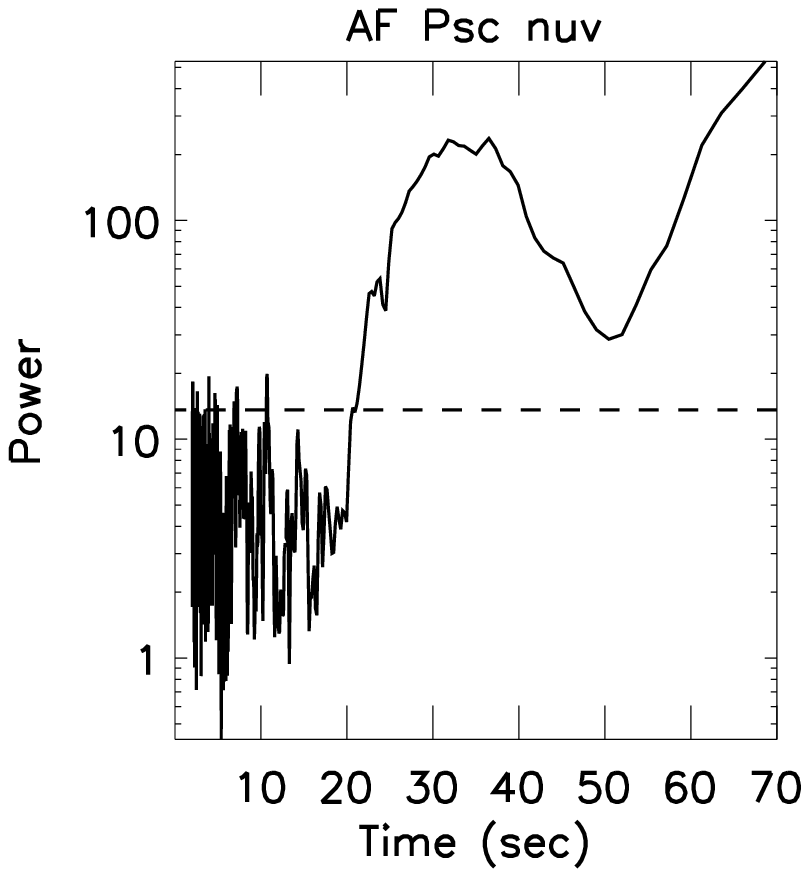}
\vspace*{-2cm}

\caption{The left two plots show the flare light-curve plus the de-trended NUV data
for AF Psc, while the central two panels show the detected oscillation periods less than
60 s plus their associated probability. The right panel shows the result of the Fourier
analysis.}
\label{GL_nuv}
\end{figure*}

\begin{figure*}[hbt]
\vspace{10.0cm}
\includegraphics{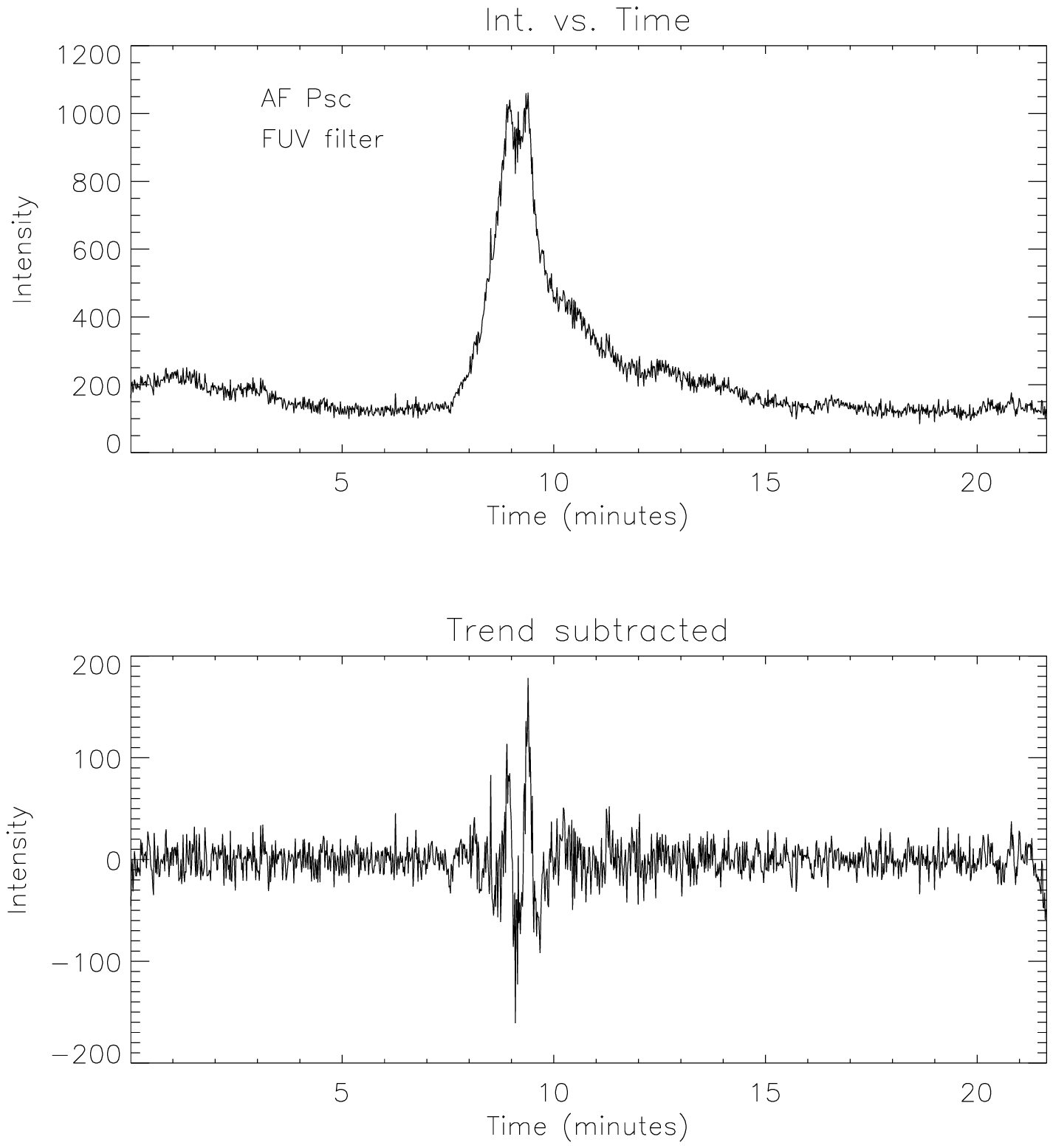}
\includegraphics{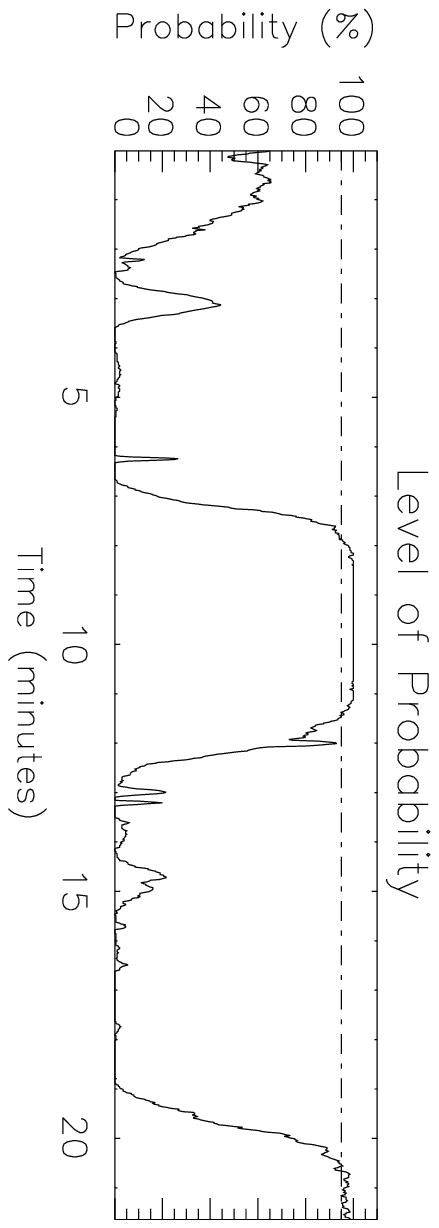}
\includegraphics{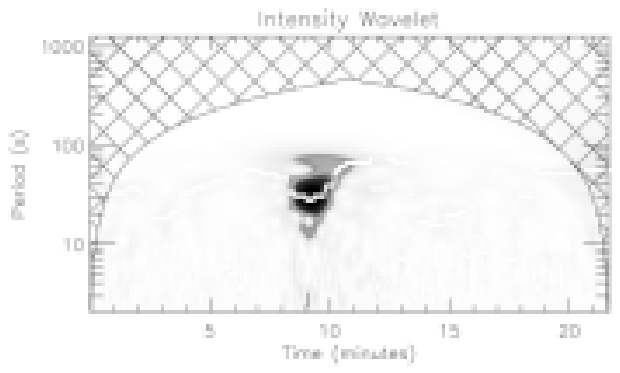}
\includegraphics{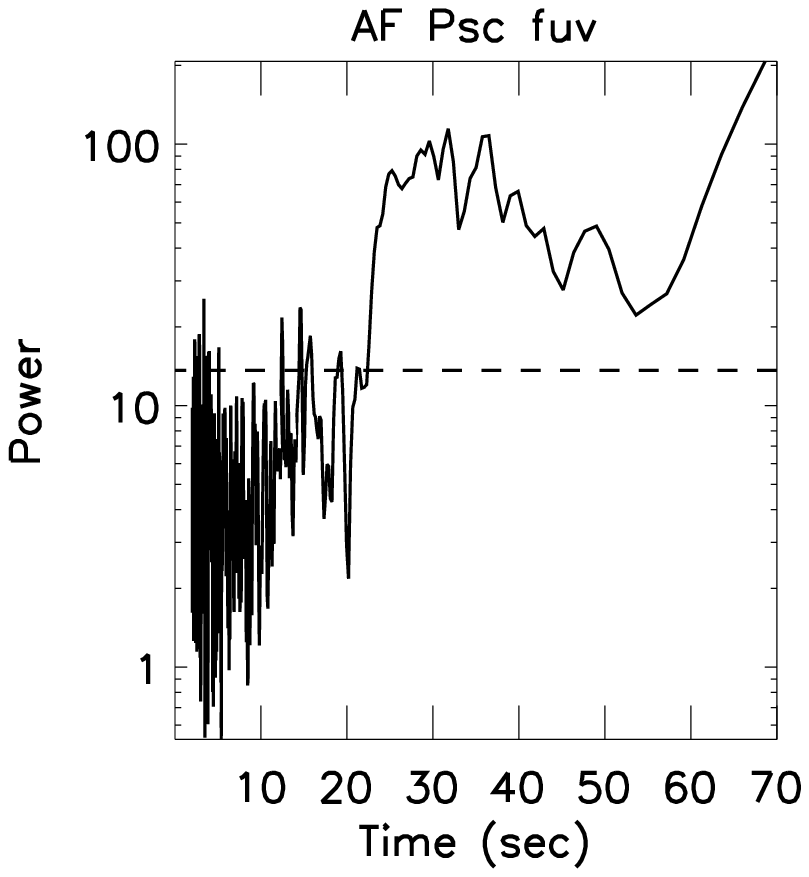}
\vspace*{-2cm}

\caption{The left two plots show the flare light-curve plus the de-trended FUV data
for AF Psc, while the central two panels show the detected oscillation periods less than
60 s plus their associated probability. The right panel shows the result of the Fourier
analysis.}
\label{GL_fuv}
\end{figure*}

\begin{figure*}[hbt]
\vspace{11.0cm}
\includegraphics{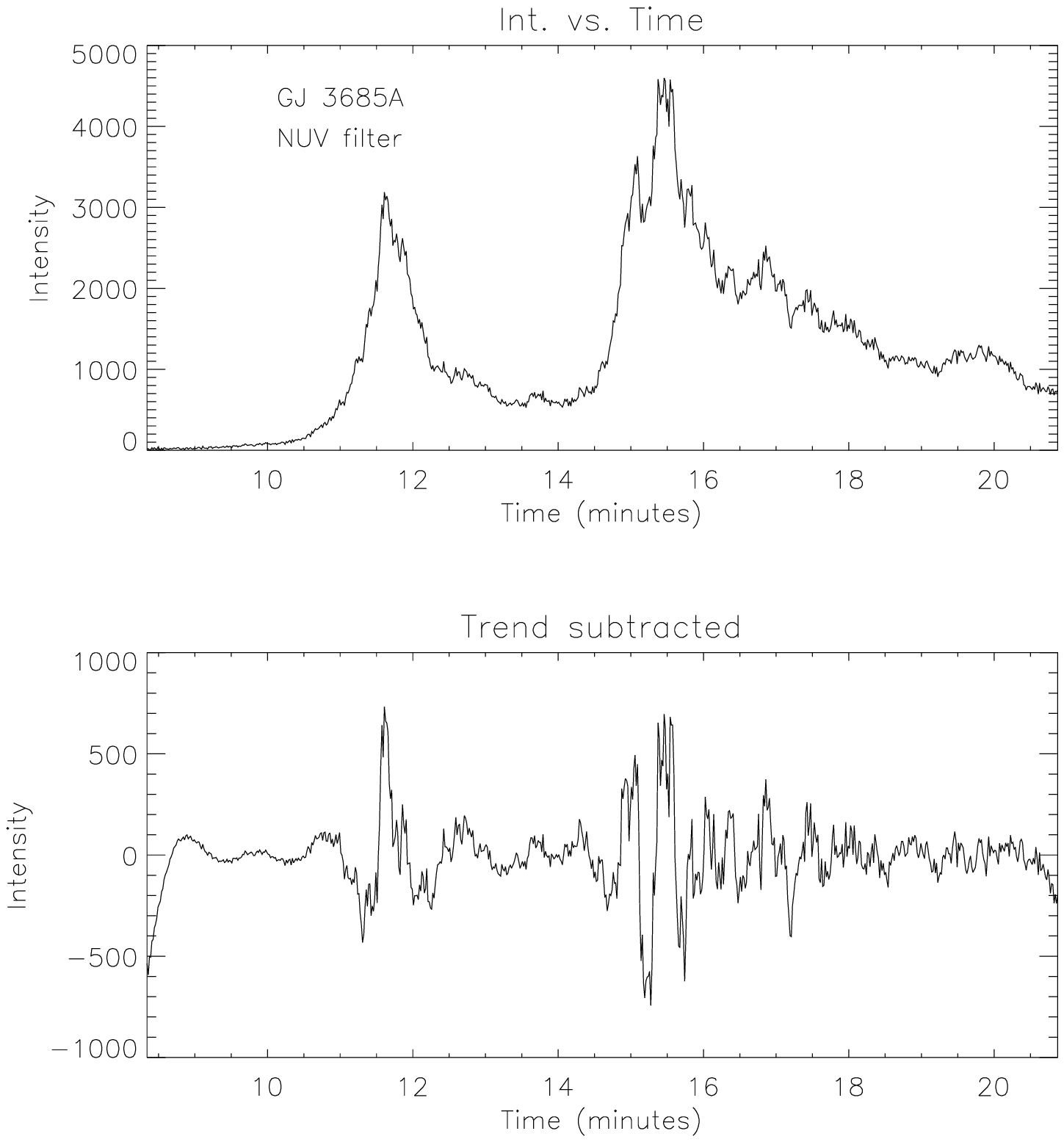}
\includegraphics{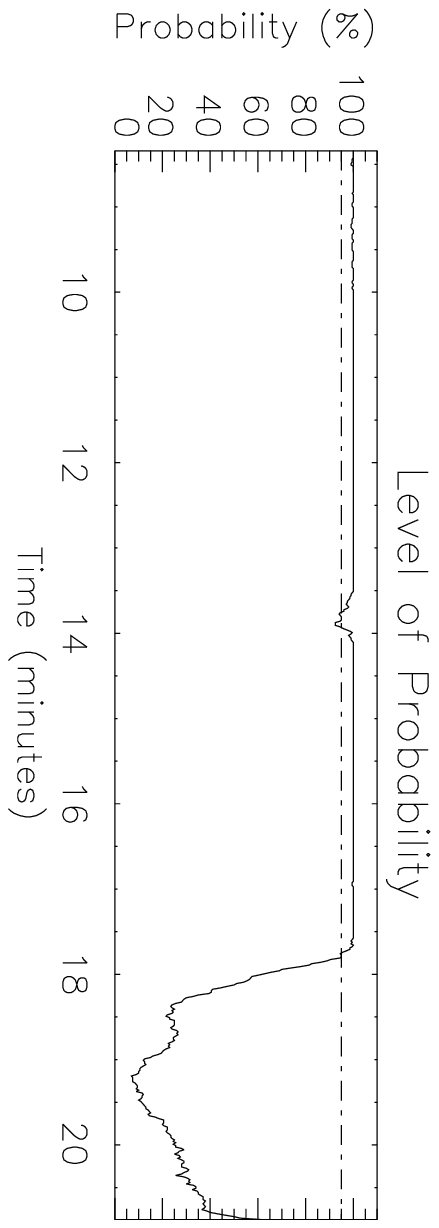}
\includegraphics{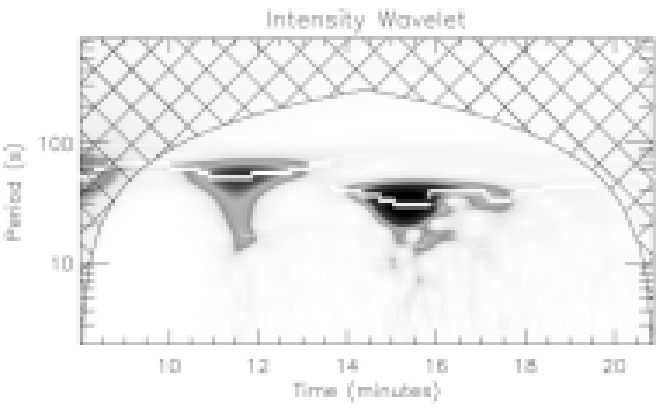}
\includegraphics{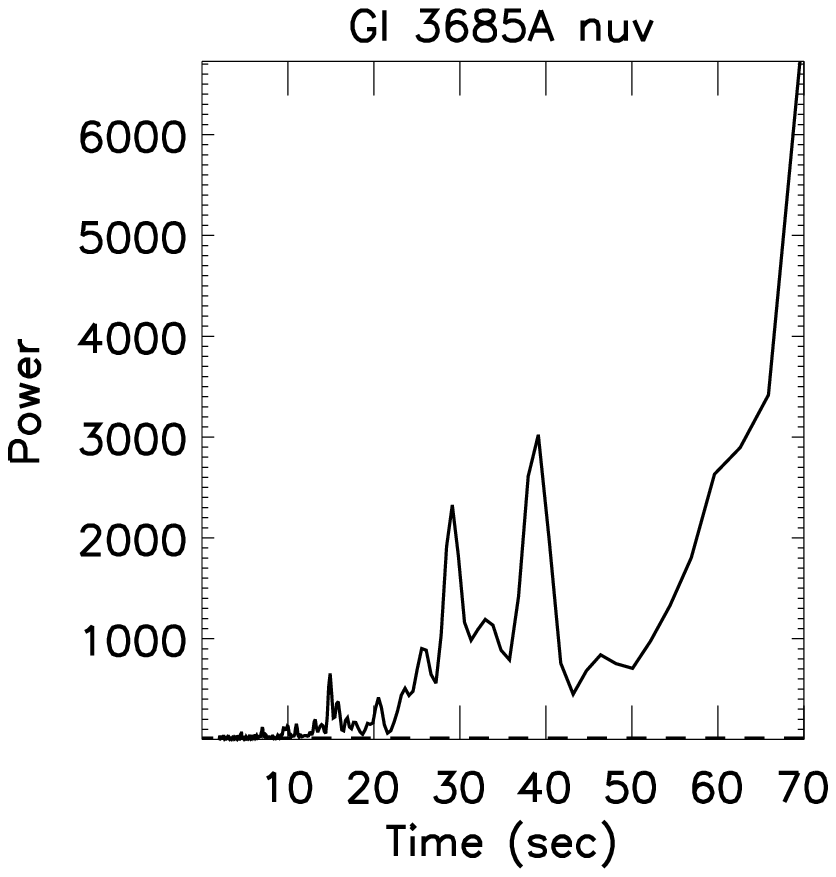}
\vspace*{-3cm}

\caption{The left two plots show the flare light-curve plus the de-trended NUV data
for GJ 3685A, while the central two panels show the detected oscillation periods less than
60 s plus their associated probability. The right panel shows the result of the Fourier
analysis.}
\label{CR}
\end{figure*}

\begin{figure*}[hbt]
\vspace{11.0cm}
\includegraphics{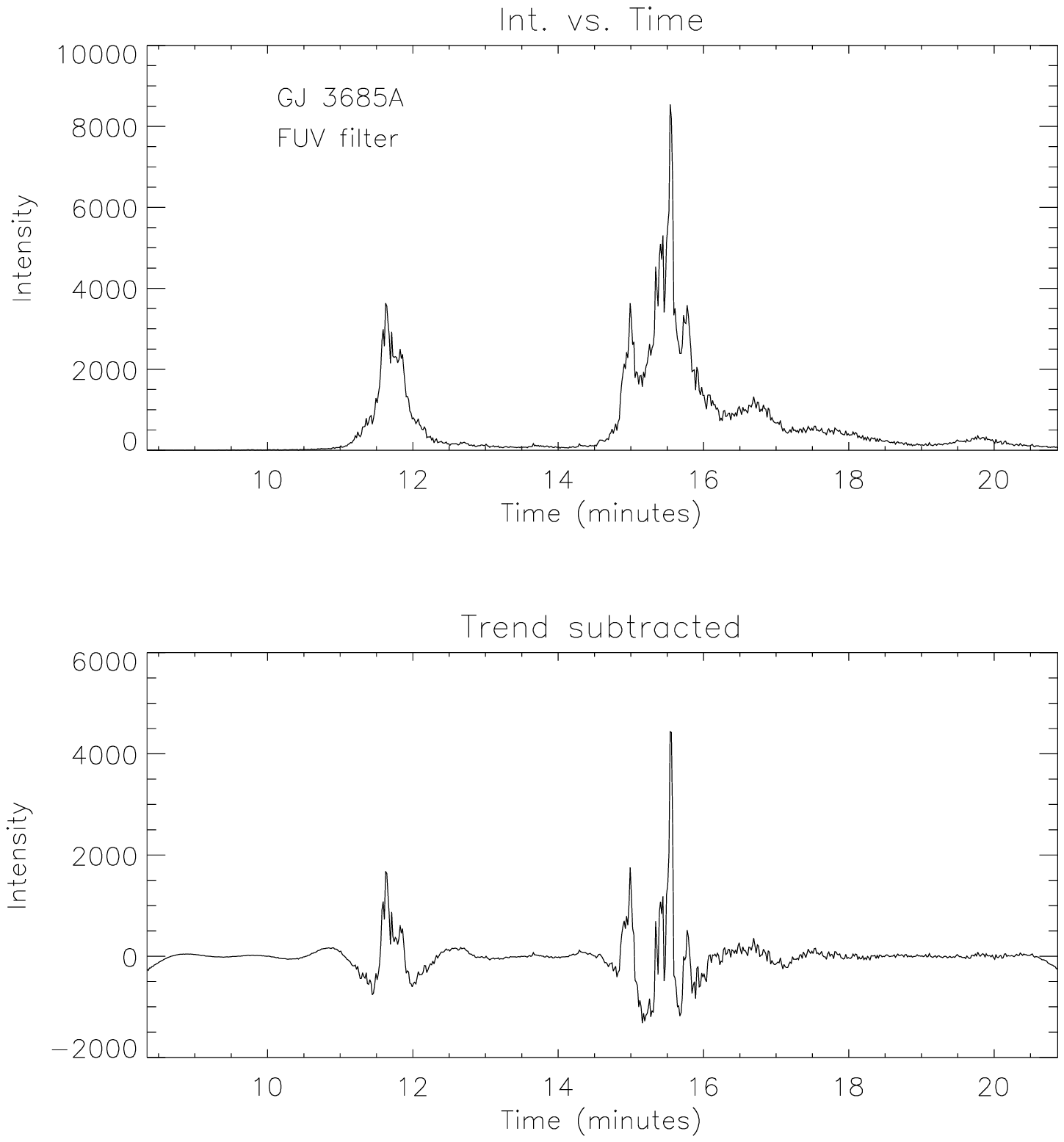}
\includegraphics{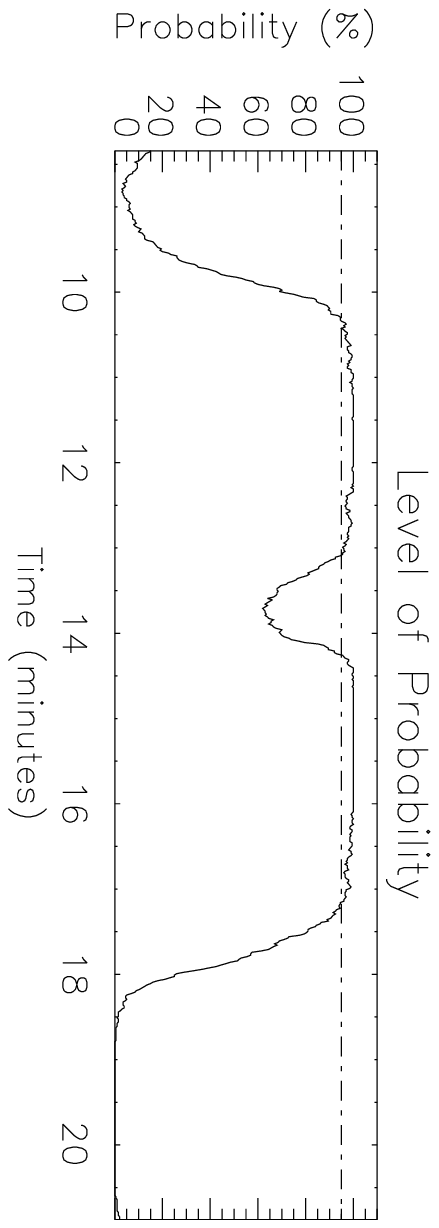}
\includegraphics{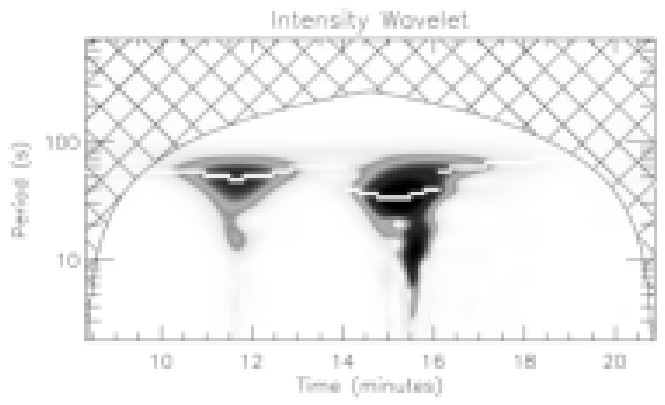}
\includegraphics{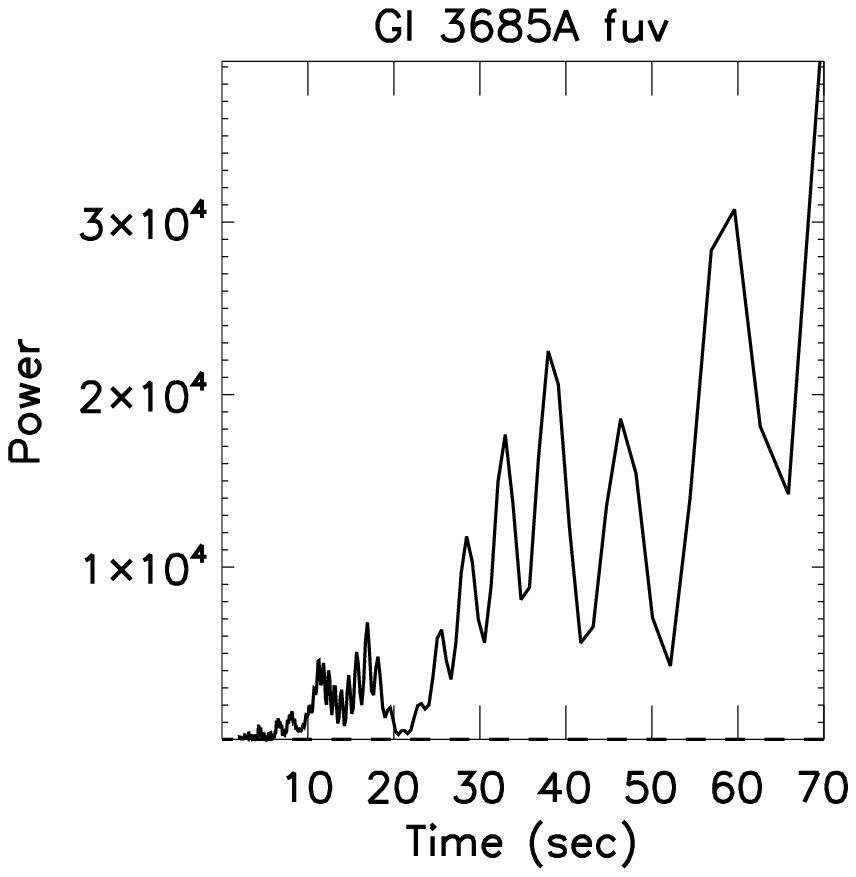}
\vspace*{-3.5cm}

\caption{The left two plots show the flare light-curve plus the de-trended FUV data
for GJ 3685A, while the central two panels show the detected oscillation periods less than
60 s plus their associated probability. The right panel shows the result of the Fourier
analysis.}
\label{sdss}
\end{figure*}

\begin{figure*}[hbt]
\vspace{11.0cm}
\includegraphics{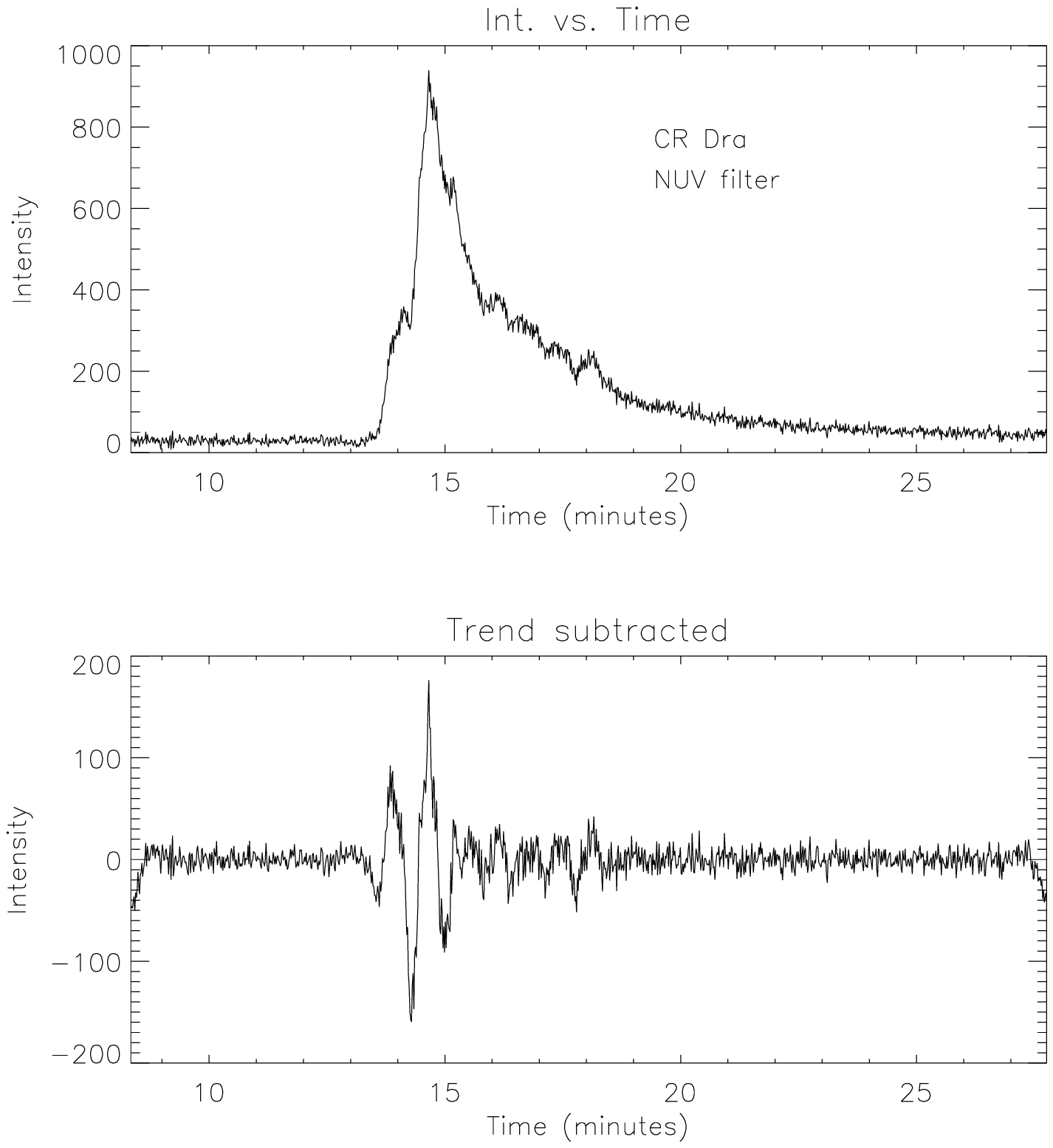}
\includegraphics{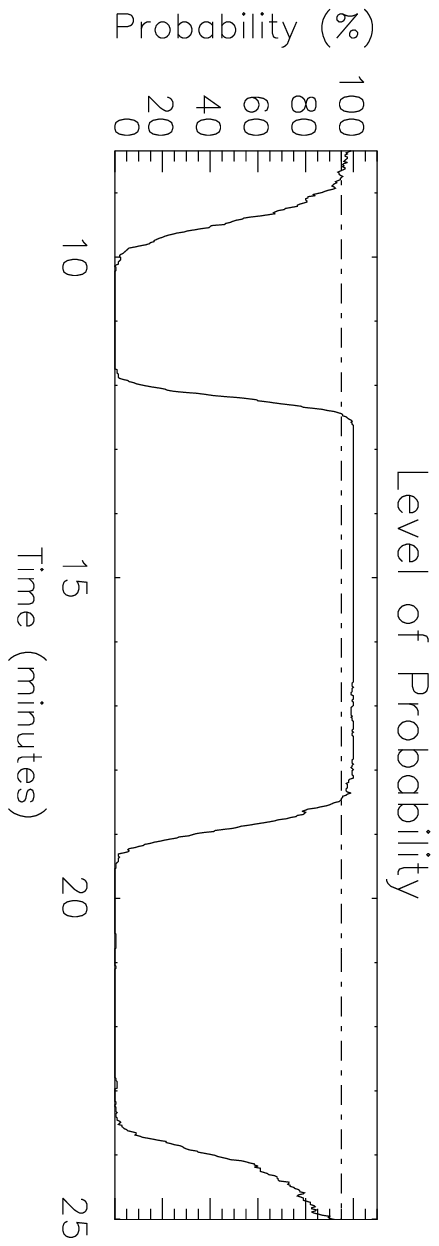}
\includegraphics{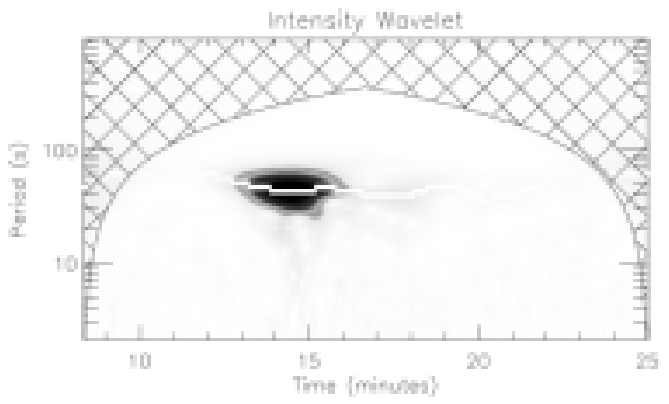}
\includegraphics{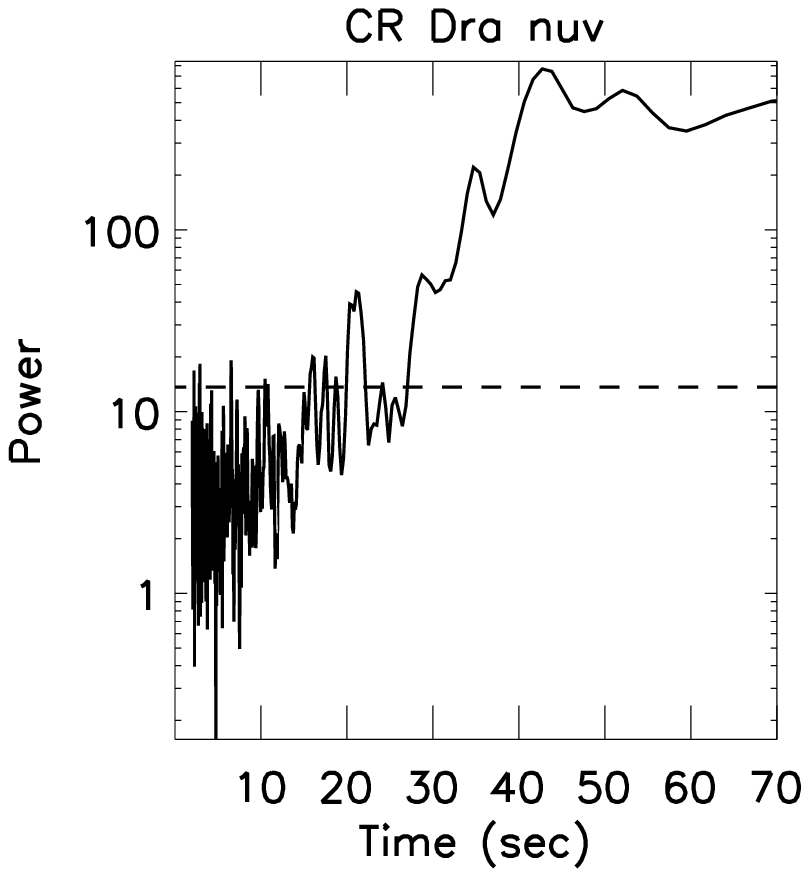}
\vspace*{-3cm}

\caption{The left two plots show the flare light-curve plus the de-trended NUV 
data for CR Dra, while the central two panels show the detected oscillation periods less than
60 s plus their associated probability. The right panel shows the result of the Fourier
analysis.}
\label{sdss}
\end{figure*}

\begin{figure*}[hbt]
\vspace{11.0cm}
\includegraphics{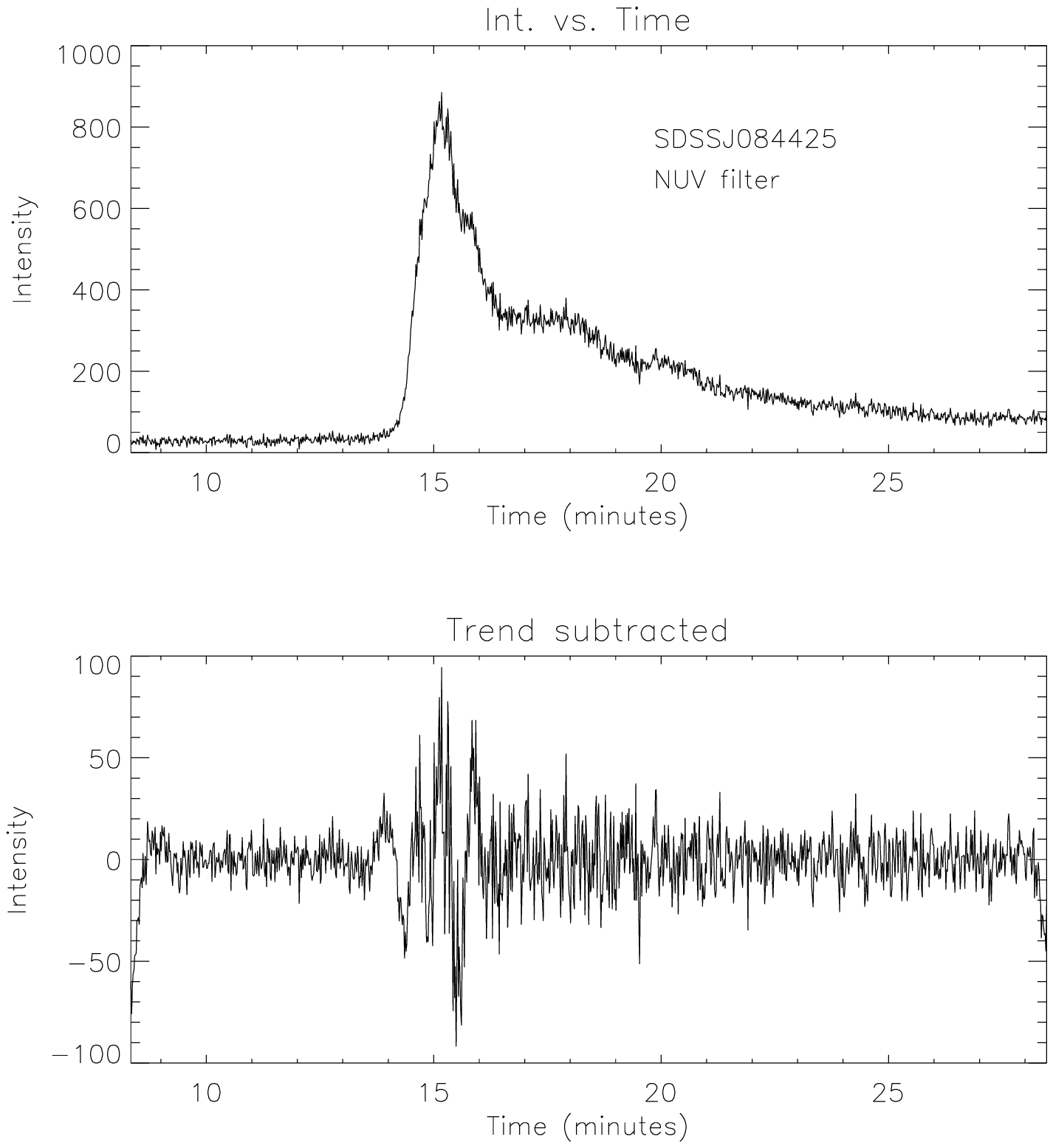}
\includegraphics{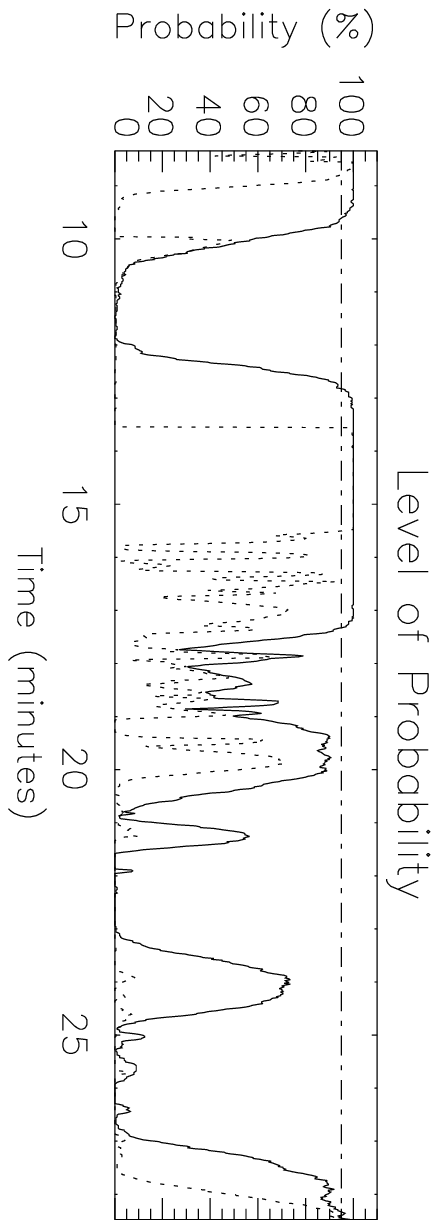}
\includegraphics{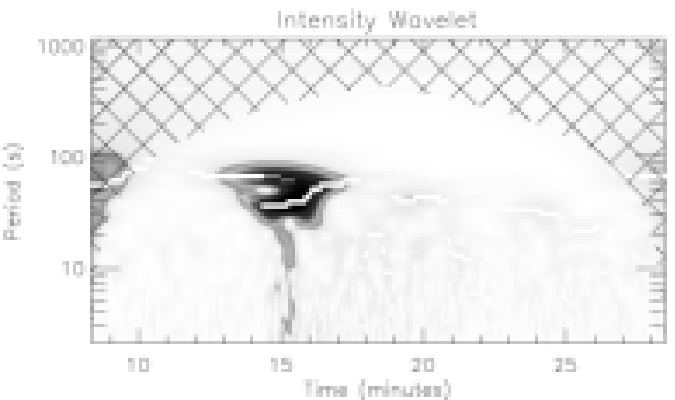}
\includegraphics{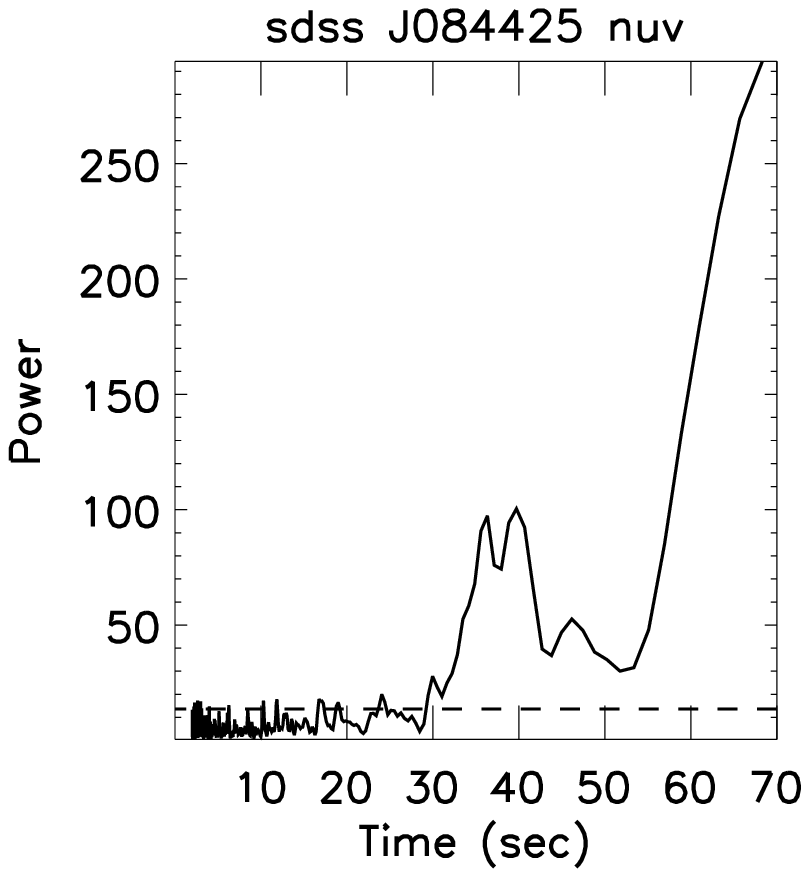}
\vspace*{-3cm}

\caption{The left two plots show the flare light-curve plus the de-trended NUV data
for SDSS J084425, while the central two panels show the detected oscillation periods less than
60 s plus their associated probability. The right panel shows the result of the Fourier
analysis.}
\label{sdss}
\end{figure*}
\begin{table}
\caption{A summary of the $\it GALEX$ FUV:NUV band flux ratios for 
the two different DEM distribution and five electron pressures as 
discussed in the text.}

\begin{tabular}{ccc}
\hline
\hline
electron pressure (cm$^{-3}$  K) & FUV:NUV & FUV:NUV\\
                                 & DEM (Q) & DEM (F) \\
\hline
$10^{15}$ & 0.59 & 1.00\\
$10^{16}$ & 0.93 & 1.66\\
$10^{17}$ & 1.27 & 2.35\\
$10^{18}$ & 1.46 & 2.54\\
$10^{19}$ & 1.84 & 3.04\\

\hline
\hline
\end{tabular}
\end{table}

\subsection{Oscillations}
Magneto-hydrodynamic oscillations associated with coronal
loops in active regions on the Sun have been widely
reported (Aschwanden et al. \cite{asch99}), but only recently have such oscillations
been observed during an X-ray flare on an active
M-dwarf star (Mitra-Kraev et al. \cite{mitra05b}).
In this latter case evidence was found for a damped oscillation
with a period of $\sim$ 750 s, from which a coronal loop length of 2.5 x 10$^{10}$ cm
was derived. Based on
this recent success in finding oscillations
in M-dwarf flare spectra, we have similarly searched for
periodicities in the observed UV photon count-rates
from the four flare events observed by $\it GALEX$ using two different analytic tools:
the wavelet and the Fourier transform methods. 
The wavelet transform is used here since it provides information on
  the location in time of where the oscillations occur, their temporal
  spread, and also the period of the oscillations. In this, the use of
  wavelets are an advantage over the more traditional Fourier power
  spectra method, which only gives information on the periods present within a
  time series, but no information on where in time that period occurs within
  the time series of data.
Details on the wavelet analysis, which 
provides information on the temporal dependence of a signal, are described in
Torrence $\&$ Compo (\cite{TC98}) \footnote{see http://atoc.colorado.edu/research/wavelets/}.
For the convolution of the time series in the  wavelet transform we have chosen
the Morlet function as defined in Torrence $\&$ Compo (\cite{TC98}). 
We note that the wavelet transform ``inspects'' the time series at 
a number of temporal scales (periods). In effect, the wavelet
transform is a bandpass filter (Torrence $\&$ Compo \cite{TC98}), and
so, by choosing an appropriate scale it is possible to filter the original time
series. This is possible since the wavelet transform is a bandpass
filter with a known response function and so it is possible to
  ``reconstruct'' the original time series using an inverse filter. In
 our case, we chose only those scales less than 60 sec and then
 reconstructed the time series containing only  periods less than
  this. The results of  applying this method can be seen in Figures 6 to 11. 
To determine whether or not any 
oscillations that were found were real, we implemented
the Linnell Nemec $\&$ Nemec (\cite{NN85}) randomization 
method that estimates the significance level of the main peak in the wavelet 
spectrum. The use of the randomization technique was performed according to
O'Shea et al. (\cite{OShea01}). Only those periods with a significance
level above 95\% were considered real and used in this work. The use
of a significance level of 95\% underlines the reliability of the
wavelet technique. To back up the use of the wavelets in detecting the oscillation
periods, we also show estimates of periods obtained from
power spectra measured using smoothed Fourier transforms. 
We used the Fourier transform
 method of Jenkins $\&$ Watts (\cite{Jenkins68}) in a
similar manner to that outlined in
Doyle et al. (\cite{Doyle99}). Significance levels of 99.9\% were
  used for the power spectra displayed in Figures 6 to 11, and shown
  as the dashed lines. The fact that the wavelet and Fourier results
  show the same periods is further evidence of the reliability
of using wavelets to measure the periods of oscillation. 
We note that Figures 6--11 show the flare light-curve, the de-trended data 
eliminating all periods greater than 60 seconds, the wavelet probability level, 
the detected periods from the wavelet and Fourier results for all 4 stellar flares. The 
wavelet probability refers to the maximum power at each time period in the wavelet 
plot, and is indicated by the over-plotted white line.

Finally, we note that using the ``post-mortem'' technique of Schwazenberg-Czerny
(\cite{Schwarz1991}) and the power spectra in Figures 6 to 11 we can 
estimate an error in the periods measured from the
power spectra of approx. 3--5 sec, which in each case is much less than the
typical periods of 30--40 sec being measured.

 Evidence for short period oscillations (less than 60s) is found
 for all of the 4 flaring sources shown in
  Figure 1. In the NUV and FUV bands of AF Psc strong oscillations with 
  a $\sim$30s period are seen in the wavelet plots. This is
  confirmed by the results from the accompanying power spectrum where 
  broad peaks at $\sim$30 sec are seen. In these plots of power spectra
 the large amount of low period (high frequency) noise can be
  easily seen. Similarly, both the NUV and
  FUV bands of GJ 3685A show evidence for short oscillation periods of $\sim$40
  sec. The main peaks in the power spectra (below 60 sec) for these
  two bands are at the same period of $\sim$40 sec. However the FUV
  band shows a more complicated structure with evidence of many more
  peaks present down to lower periods than that seen in the NUV band. 
 For the NUV data of GJ 3685A, the wavelet analysis suggest periods 
of around 30 sec during the initial stages of the second flare, increasing 
to 40 sec during the decay phase. This is also seen in the Fourier
  results.
Similarly, the power spectrum CR Dra NUV shows strong evidence for a 
period around 40 sec. In
  the wavelet spectrum plot for this source the oscillation is shown
  as a single peak at a period of $\sim$47 sec, lasting for only
  the time period between 13 and 16 minutes. The failure of the wavelet
  spectrum to resolve two peaks at $\sim$40 and 50 seconds is
  related to its relatively poorer temporal resolution compared to the
  Fourier power spectrum, i.e., the restricted number of scales that
  can be chosen. This is an inherent drawback of the wavelet method
  that is compensated by the information the wavelet gives us on where
  an oscillation is occurring in time during the observing period.
For the source SDSS J084425, in the NUV we again find a period of
 $\sim$40 sec, measured from both the Fourier and wavelet methods. 

Several authors have used radiative and conductive cooling times (e.g., Hawley 
et al. \cite{hawley95}) to estimate flare loop lengths. During the rise phase, strong 
evaporative heating is dominant while during the decay phase radiative cooling 
is dominant. At flare maximum, the above authors equated these two quantiles, 
giving the loop length as 
\begin{equation}\label{rad}
L = \frac{1500}{(1 - x_d^{1.58})^{4/7}} \, \tau_d^{4/7} \, \tau_r^{3/7} \, \surd T
\end{equation}
where $\tau_d$ is the flare decay time, $\tau_r$ is the flare rise time, T the
apex flare temperature, and $x_d = \surd (c_d/c_{max})$, with $c_{max}$ the 
peak flare count rate and $c_d$ the count rate at the end of the flare. For the 
present flares, the typical rise time is $\sim$ 70 sec, the decay time 
$\sim$ 300 sec and $c_d/c_{max} \sim$ 0.1. Unfortunately, we do not know 
the plasma temperature, but assuming a range of 
temperatures from $5 \times 10^6$ K to $20 \times 10^6$~K, implies loop 
lengths of $0.5 - 1 \times 10^9$~cm. This temperature range is in broad agreement 
with the results from observations, e.g. in the modelling of X-ray data, 
Sciortino et al. (\cite{Sciortino99}) derived a maximum temperature of $20 \times 10^6$~K for 
EV Lac and a much higher temperature of $300 \times 10^6$~K for AD Leo. 
Several authors have also 
modelled the non-flare X-ray spectral data, e.g. Osten et al.(\cite{Osten05}) and 
Sciortino et al. (\cite{Sciortino99}) used two and even a three-component model to fit their 
data. The implied loop lengths from their work was 0.1R$_{*}$ for loop 
temperatures around a few millions degrees, with even smaller loops at the 
higher temperatures. 

Based on the spectral analysis outlined in the previous section, the implied
electron pressure during a flare may be close to $10^{18} - 10^{19}$ cm$^{-3}$ K, implying
large electron densities, consistent with measurements.  
Emslie (\cite{Emslie81}) has suggested that quasi-periodic increases in the hard X-ray flux 
from solar flares may be the result of the local gas pressure attaining very 
large values, significantly larger than the values suggested in the previous 
section. However, the idea that transient increases in the electron density 
result in quasi-periodic increases in the line flux is never-the-less valid. 
The quasi-periodic increases in the line flux could be interpreted in terms of 
fundamental mode quasi-standing waves, these being the easiest to excite. 
Such interpretation has already been given 
by Mitra-Kraev et~al. (\cite{mitra05b}) to flare oscillations on an active M-dwarf 
star. The interpretation in terms of propagating waves or higher harmonics is less 
plausible given the 
spatial resolution of the observations and the cancellation effect along the loop, i.e.,
any density increase on one side of the loop would be accompanied by a corresponding
decrease one the other side. The theory of coronal loop oscillations was developed by
Roberts et~al. (\cite{roberts84}). The periods of the oscillations are given by the expression
\begin{equation}
P=\frac{2L}{c},
\end{equation}
where $c$ corresponds to either the fast kink mode, the fast sausage mode, or the slow sausage
mode. The fast kink mode is almost incompressible and can therefore be discarded from further 
discussion. According to Roberts et~al. (\cite{roberts84}), the wavenumbers required for the 
existence of the fast sausage wave are high under coronal conditions due to the cutoff property.
In the case of a standing mode, the wavenumber is determined by the loop dimensions and is
below the cutoff value. The remaining possibility is the slow sausage mode. The relationship 
between the period P, loop length L and temperature T for the fundamental mode
slow standing wave is approximately given by
\begin{equation}
L \approx \frac{P}{2} c_T \,\,\, \mbox{or} \,\,\, L= 7600 P c_{Ts}\surd T ,
\end{equation}
with $c_{Ts}$ being the ratio between the cusp speed $c_T$ and the sound speed $c_s$. Comparing 
the above relationship with formula~(\ref{rad}) we find a ratio $c_{Ts}=0.73$ for a
period of $40$~s. In comparison, Mitra-Kraev et~al. (\cite{mitra05b}) found a ratio
$c_{Ts}=0.82$. This means that the plasma $\beta$ inside the loop (i.e., the ratio between 
the thermal pressure and the magnetic pressure) is slightly higher in the present case. 

Slow sausage (acoustic) fundamental mode standing waves have been recently detected on the Sun
(Wang et~al. \cite{wang03}). These waves are caused by the injection of hot chromospheric 
plasma at the footpoints of the loops. Taroyan et~al. (\cite{taroyan05}) showed that such 
oscillations could be readily excited if the duration of the plasma injection at the loop
footpoint matches the natural fundamental mode period of the loop. The closeness of these two  
quantities would result in quasi-standing acoustic oscillations. In the present case, the 
injection of plasma into the loop could be associated with the flare and should last 
approximately 40~s. {\bf Finally, on a precautionary note, we acknowledge that the sequential heating
of evenly spaced loops
at a constant rate (i.e. in an arcade) could also reproduce the oscillation
signature that we have found in these data. We refer the reader to the excellent
review of coronal oscillations by Roberts (\cite{roberts00}) for further details of this,
and other possible oscillation mechanisms.}

\section{Conclusion}
We have presented near and far ultraviolet light-curves for stellar flares observed
with the NASA $\it GALEX$ satellite
on the four nearby dMe-type stars GJ 3685A, CR Dra, AF Psc and
SDSS J084425. These data were recorded as time-tagged UV photon events with
a temporal resolution of $<$ 0.01 seconds, thus enabling a detailed time-analysis
of both their quiescent and flaring states. During 700 seconds of quiescence,
prior to the flare outbursts on both CR Dra and SDSS J084425, no statistical evidence
was found for micro-flare outbursts in the NUV with data binned in
0.2, 1.0 and 10.0 second time-intervals.

We have used a modified differential emission model (DEM) curve
of the flare star EV Lac in conjunction
with the CHIANTI atomic database to produce theoretical  DEM curves for a typical
dMe star in both quiescence and a flaring state. From the quiescent DEM distribution
we were able to reproduce FUV:NUV flux ratios greater
than unity, consistent with the measurements during
the flares. A critical input parameter in determining the FUV:NUV
ratio is the value of plasma electron density during the flare event. It was also found that the 
major single emitter in the FUV filter is C~{\sc iv}, although weaker emission lines and the
continuum provide $\sim$50\% of the flux. In the NUV filter, a significant contribution 
comes from upper chromosphere/lower transition region emission lines, in 
addition to continuum emission. Several coronal lines are present in both 
filters, with $\sim$10\% contribution in the NUV filter, whose main contributer is
from continuum processes.

We have also searched for periodicities in the observed UV count-rate data
recorded during the four flare events to search for coronal loop oscillations.
Using both the wavelet and Fourier transform analysis methods on these flare data,
we have found compelling statistically significant evidence for oscillations in both
of the NUV and FUV data
for flares on AF Psc and GJ 3685A, and in the NUV data for the
flares on CR Dra and
SDSS J084425 with periodicities in 30-40 second range. Assuming typical 
values of electron pressure during a flare, these periodicities may be due to
acoustic waves in coronal loops of length $\sim$ $10{^9}$ cm for a plasma 
with a temperature range of $5-20 \times 10^6$ K. These values suggest
a loop length of less than $1/10^{th}$ of the M-dwarf stellar radii.
Finally we note that although oscillations
have been previously observed on the Sun and in X-ray observations of the
M-type dwarf AT Mic, we believe that this is the first detection of non-solar 
coronal loop flare oscillations at ultraviolet wavelengths.

\begin{acknowledgements}
We gratefully acknowledge NASA's support for construction, operation,
and science analysis for the GALEX mission,
developed in cooperation with the Centre National d'Etudes Spatiales
of France and the Korean Ministry of
Science and Technology. We acknowledge the dedicated
team of engineers, technicians, and administrative staff from JPL/Caltech,
Orbital Sciences Corporation, University
of California, Berkeley, Laboratoire d'Astrophysique de Marseille,
and the other institutions who made this mission possible.
Financial support for this research was provided by the
NASA $\it GALEX$ Guest Investigator program, administered by
the Goddard Spaceflight Center in Greenbelt, Maryland.
This publication makes use of data products from the SIMBAD database,
operated at CDS, Strasbourg, France. CHIANTI is a collaborative project
involving NRL (USA), RAL (UK) and the University of Florence (Italy) and
the University of Cambridge (UK).

\end{acknowledgements}

\end{document}